\documentclass[]{aa}
\usepackage{graphics}
\usepackage{txfonts}
\begin{document}
\title{Millimeter observations of Planetary Nebulae:}
 \subtitle{a contribution to the Planck pre-launch catalogue}
\author{
G. Umana\inst{1},
P. Leto\inst{1,}\inst{2}
C. Trigilio\inst{1},
C. S. Buemi\inst{1},
P. Manzitto\inst{3},
S. Toscano\inst{3},
S. Dolei\inst{3},
L. Cerrigone\inst{3,}\inst{4}
}

\offprints{G. Umana \email{Grazia.Umana@oact.inaf.it}}

\institute{
INAF - Osservatorio Astrofisico di Catania, Via S. Sofia 78, 95123 Catania, Italy
\and
INAF Istituto di Radioastronomia, Sezione di Noto, C.P. 161, Noto(SR), Italy
\and 
Universit\'a  di Catania, Dipartimento di Fisica e Astronomia, Via S. Sofia 78 , 95123 Catania, Italy
\and 
Harvard-Smithsonian Center for Astrophysics, Cambridge, MA 02138, USA
}

\date{Received; Accepted}

\titlerunning{Millimetre observations of PNe}
\authorrunning{G. Umana et al.}

\abstract  
{} 
{We present new millimetre 43 GHz observations of a sample of radio-bright Planetary Nebulae.
Such observations were carried out to have a good determination of the high-frequency radio spectra of the sample in order to evaluate, together with far-IR measurements (IRAS), the fluxes emitted by the selected source in the millimetre and  sub-millimetre band. This spectral range,  even very important to constraint the physics of circumstellar environment, 
is still far to be completely exploited.}
{To estimate the millimetre and sub-millimetre fluxes, we extrapolated and summed together the ionized gas  (free-free radio emission) and dust (thermal emission) contributions at this frequency range. By comparison of the derived flux densities to the foreseen sensitivity we investigate the possible detection of such source for all the channels of the forthcoming ESA's PLANCK mission.}
{We conclude that almost 80\% of our sample will be detected by PLANCK, with the higher detection rate in the higher frequency channels, where there is a good combination of brighter intrinsic flux from the sources and reduced extended Galactic foregrounds contamination despite a worst instrumental sensitivity.
From the new 43 GHz, combined with single-dish 5 GHz observations from the literature, we derive radio spectral indexes, which are consistent with optically thin free-free nebula. This result indicates that the high frequency radio spectrum of our sample sources is dominated by thermal free-free and other emission, if present, are negligible.  
}
{} 


     \maketitle
%

\section{Introduction}

The PLANCK ESA mission will provide us with  nearly full sky maps
over a wide range of frequencies, from 30 to 900 GHz. Therefore, even if designed  for cosmological 
studies, the mission  will have profund impact on fundamental Physics and Galactic and extragalactic astrophysics.
Planck will be sensitive to the millimetre emission from dusty envelopes of stars and we expect the 
dusty circumstellar envelopes, that characterize the latest stages of stellar evolution, to be source of
relevant foreground contamination. 
When a low or intermediate mass star is approaching the end of its
evolution, it goes through a period of heavy mass-loss known as Asymptotic 
Giant Branch (AGB) phase. The ejected envelopes are partially condensed
in dust grains and completely obscure the central star.
Immediately after the AGB evolutionary phase, the mass-loss stops and the
stars may become optically visible as the dusty shells disperse (Proto Planetary
Nebula, PPN phase).
Eventually, once it reaches a temperature of 2 -- 3 $\times 10 ^4$ K, the central star
starts to ionize the AGB shell and a Planetary Nebula will form.
Dusty envelopes re-radiate the absorbed stellar light showing a clear
signature in the far-infrared spectrum, i.e. an IR excess with peculiar
IRAS colours. In addition to that, PNe show  a radio continuum  
due to free-free emission from the fraction of the CSE ionized by the central star.

\begin{table*}
 \caption{The selected sample}
 \label{tab_posizioni}
 \begin{center}

 \vspace{0.3cm}
 
 \begin{tabular}{cccc|cccc}
 \hline
 \hline
 IAU Name    & Other Name   & R.A.\ (J2000) &Dec. \ (J2000)    &IAU Name &
 Other Name  & R.A.\ (J2000)  &   Dec. \ (J2000)  \\
 PN G&& [\ h\ m\ s] &[$^{\circ}\ ^{\prime}\ ^{\prime\prime}$]  &PN G&&[\ h\
 m\ s] &[$^{\circ}\ ^{\prime}\ ^{\prime\prime}$] \\
 \hline
 $000.3+12.2$ &IC 4634         & 17 01 33.6    & $-21$ 49 33.1   
 &$093.4+05.4$ &NGC 7008         & 21 00 32.7      & $+54$ 32 39.4   \\
 $002.4+05.8$ &NGC 6369        & 17 29 20.5    & $-23$ 45 35.0   
 &$093.5+01.4$ &PN M 1-78        & 21 20 44.8      & $+51$ 53 27.5   \\
 $003.1+02.9$ &PN Hb 4         & 17 41 52.8    & $-24$ 42 09.3   
 &$096.4+29.9$ &NGC 6543         & 17 58 33.4      & $+66$ 37 58.8   \\
 $006.7-02.2$ &PN M 1-41       & 18 09 30.6    & $-24$ 12 28.7   
 &$097.5+03.1$ &PN A66 77        & 21 32 10.2      & $+55$ 52 43.2   \\
 $007.2+01.8$ &PN HB 6         & 17 55 07.0    & $-21$ 44 41.0   
 &$106.5-17.6$ &NGC 7662         & 23 25 53.9      & $+42$ 32 04.7   \\
 $008.0+03.9$ &NGC 6445        & 17 49 15.0    & $-20$ 00 33.7   
 &$107.8+02.3$ &NGC 7354         & 22 40 19.9      & $+61$ 17 08.0   \\
 $008.3-01.1$ &PN M 1-40       & 18 08 26.0    & $-22$ 16 53.4   
 &$120.0+09.8$ &NGC 40           & 00 13 01.0      & $+72$ 31 19.6   \\
 $009.4-05.0$ &NGC 6629        & 18 25 42.5    & $-23$ 12 11.3   
 &$130.9-10.5$ &NGC 650-51       & 01 42 19.7      & $+51$ 34 31.7   \\
 $009.6+14.8$ &NGC 6309        & 17 14 04.3    & $-12$ 54 37.2   
 &$138.8+02.8$ &IC 289           & 03 10 19.3      & $+61$ 19 00.4   \\
 $010.1+00.7$ &NGC 6537        & 18 05 13.1    & $-19$ 50 34.4   
 &$144.5+06.5$ &NGC 1501         & 04 06 59.3      & $+60$ 55 14.7   \\
 $010.8-01.8$ &NGC 6578        & 18 16 16.5    & $-20$ 27 03.4   
 &$165.5-15.2$ &NGC 1514         & 04 09 16.9      & $+30$ 46 32.0   \\
 $011.7-00.6$ &NGC 6567        & 18 13 45.2    & $-19$ 04 35.6   
 &$166.1+10.4$ &IC 2149          & 05 56 23.9      & $+46$ 06 17.4   \\
 $020.9-01.1$ &PN M 1-51       & 18 33 29.0    & $-11$ 07 26.3   
 &$173.7+02.7$ &PP 40            & 05 40 52.7      & $+35$ 42 18.6   \\
 $025.8-17.9$ &NGC 6818        & 19 43 57.8    & $-14$ 09 11.8   
 &$194.2+02.5$ &J 900            & 06 25 57.3      & $+17$ 47 27.6   \\
 $027.7+00.7$ &PN M 2-45       & 18 39 21.9    & $-04$ 19 52.6   
 &$197.8+17.3$ &NGC 2392         & 07 29 10.8      & $+20$ 54 41.6   \\
 $033.8-02.6$ &NGC 6741        & 19 02 37.0    & $-00$ 26 57.2   
 &$206.4-40.5$ &NGC 1535         & 04 14 15.8      & $-12$ 44 22.3   \\
 $034.6+11.8$ &NGC 6572        & 18 12 06.3    & $+06$ 51 12.4   
 &$215.2-24.2$ &IC 418           & 05 27 28.2      & $-12$ 41 50.2   \\
 $035.1-00.7$ &PN Ap 2-1       & 18 58 10.5    & $+01$ 36 57.5   
 &$221.3-12.3$ &IC 2165          & 06 21 42.8      & $-12$ 59 13.9   \\
 $037.7-34.5$ &NGC 7009        & 21 04 10.8    & $-11$ 21 48.5   
 &$234.8+02.4$ &NGC 2440         & 07 41 55.4      & $-18$ 12 30.5   \\
 $039.8+02.1$ &PN K 3-17       & 18 56 18.2    & $+07$ 07 26.2   
 &$254.6+00.2$ &NGC 2579         & 08 20 54.1      & $-36$ 13 00.0   \\
 $041.8-02.9$ &NGC 6781        & 19 18 28.1    & $+06$ 32 20.0   
 &$258.1-00.3$ &Hen 2-9          & 08 28 28.0      & $-39$ 23 39.4   \\
 $043.1+37.7$ &NGC 6210        & 16 44 29.5    & $+23$ 47 59.9   
 &$259.1+00.9$ &Hen 2-11         & 08 37 08.1      & $-39$ 25 04.9   \\
 $045.7-04.5$ &NGC 6804        & 19 31 35.1    & $+09$ 13 30.2   
 &$261.0+32.0$ &NGC 3242         & 10 24 46.1      & $-18$ 38 32.3   \\
 $050.1+03.3$ &PN M 1-67       & 19 11 31.1    & $+16$ 51 32.0   
 &$294.1+43.6$ &NGC 4361         & 12 24 30.8      & $-18$ 47 04.0   \\
 $054.1-12.1$ &NGC 6891        & 20 15 08.9    & $+12$ 42 15.4   
 &$342.1+10.8$ &NGC 6072         & 16 12 58.4      & $-36$ 13 46.6   \\
 $063.1+13.9$ &NGC 6720        & 18 53 35.1    & $+33$ 01 45.1   
 &$349.5+01.0$ &NGC 6302         & 17 13 44.5      & $-37$ 06 11.6   \\
 $064.7+05.0$ &BD+30 3639      & 19 34 45.2    & $+30$ 30 59.2   
 &$352.6+00.1$ &PN H 1-12        & 17 26 24.3      & $-35$ 01 41.8   \\
 $082.1+07.0$ &NGC 6884        & 20 10 23.7    & $+46$ 27 40.0   
 &$352.8-00.2$ &PN H 1-13        & 17 28 27.7      & $-35$ 07 30.4   \\
 $083.5+12.7$ &NGC 6826        & 19 44 48.2    & $+50$ 31 31.3   
 &$358.5+02.6$ &PN HDW 8         & 17 31 47.3      & $-28$ 42 03.5   \\
 $086.5-08.8$ &PN Hu 1-2       & 21 33 08.2    & $+39$ 38 08.3   
 &$358.5+05.4$ &PN M 3-39        & 17 21 11.5      & $-27$ 11 37.0   \\
 $089.0+00.3$ &NGC 7026        & 21 06 18.7    & $+47$ 51 07.5   
 &$359.3-00.9$ &Pn HB 5          & 17 47 56.3      & $-29$ 59 40.6   \\
 \hline
 \end{tabular}
 \end{center}
 \end{table*}

From a feasibility study, Umana et al. (\cite{Umana06}) concluded that a sizable 
($\approx 300$) sample of AGB and post-AGB stars would be detected during the mission and derived 
estimates for the expected flux densities at various Planck channels.
However, the simulations carried out by Umana et al. (\cite{Umana06}), on a sub-sample of PNe rely only on NVSS fluxes, obtained at 1.4 GHz, extrapolated to the Planck frequencies.  
This leads to underestimate the contribution due to free-free
as, at this frequency, PNe are often optically thick (Siodmiank \& Tylenda \cite{st01},  Luo et al. \cite{luo_etal05}).

 PNe  are among the brightest Galactic radio sources. Some of them could also reach a flux density of Jy level.
 More than 800 have been  detected at least at one frequency  and for 200
 morphological and spectral information (between 1.4 and 22 GHz) were obtained.
 Most of the higher frequency (22 GHz) data were obtained with interferometers (i.e. VLA) while
 very little single-dish, high frequency measurements are available.

 In this paper we present new 43 GHz, single-dish observations by using the 32 m INAF-IRA 
 Radiotelescope at Noto of a sample of PNe, which are potential foregrounds for PLANCK.
 The main goal of this project is to obtain reliable estimates of flux density expected at PLANCK channels,
 by building  and modeling their spectral energy distribution (SED). This in turn will contribute to the
 compilation of the PLANCK pre-launch catalogue. We stress here that 43 GHz is one of the observing channels of the forthcoming PLANCK mission. Therefore, at this frequency band, we would obtain a direct measurement of the expected flux and not an extrapolation.
 
  As added value, the present work provides the first sizeable dataset of 43 GHz measurements 
  of PNe, that constitute strong  constraints to the observed SEDs in the very important spectral region where free-free emission and thermal dust emission may overlap.  While interferometric high frequency observations provide us with detailed morphological information they  
 quite often fail to entirely recover the extended emission. This eventually leads to underestimate the total radio flux density. This problem is  overcame by single-dish observations.
  Assesting the SEDs in the radio-millimetre spectral range is a crucial step 
  for the study of the physics of dusty envelopes around PN. A correct evaluation of the free-free contribution, up to millimetre 
  range, when combined with information provide by far-IR observations, would allow to determine the 
  presence of an excess due to the presence of a cold dust component/s 
(Gee et al. \cite{gee_etal84}; Hoare et al. \cite{hoare_etal92}; Kemper et al. \cite{kemper_etal02}) 
or of alternative emission mechanisms 
  (Casassus et al. \cite{cas_etal07}).
  Since PNe and their progenitors are believed to be among the major sources of recycled
 interstellar matter,  determining the properties of the dust ejected in the ISM is  very important to the study of the Galaxy
 evolution in general.

\begin{table*}
\begin{center}
\caption{Results}
\label{tab_misure}
\begin{tabular}{crrclr|crrclr}
\hline
\hline
IAU Name     & $S_\mathrm{43 ~GHz}$ & $\sigma_\mathrm{43 ~GHz}$  &$\theta_\mathrm{1.4 ~GHz}$ & Ref.$^{\ast}$ &$S_\mathrm{c~43 ~GHz}$ &IAU Name  &$S_\mathrm{43 ~GHz}$
& $\sigma_\mathrm{43 ~GHz}$  &$\theta_\mathrm{1.4 ~GHz}$ & Ref.$^{\ast}$ &$S_\mathrm{c~43 ~GHz}$ \\
PN G         & [mJy]   & [mJy]  & [arcsec]  &  & [mJy] &PN G          & [mJy]   & [mJy]  & [arcsec]  &  & [mJy]    \\
\hline                                                                                  
$000.3+12.2$ & $<$ 180    & 60  & 10.0  & C H     &      &$093.4+05.4$  & $<$ 240    & 80   & 49.2    & A E   &      \\
$002.4+05.8$ & 1330       & 110 & 20.7  & C H     &1530  &$093.5+01.4$  & 560        & 35   & 14.6    & A E   &600   \\
$003.1+02.9$ & 100        & 20  & 11.6  & C H     &105   &$096.4+29.9$  & 400        & 70   & 11.9    & A E   &420   \\
$006.7-02.2$ & 300        & 90  & 21.0  &         &345   &$097.5+03.1$  & 230        & 25   & 33.4    & A E   &320   \\
$007.2+01.8$ & 310        & 50  & 10.6  & C H     &320   &$106.5-17.6$  & 610        & 40   & 12.0    & A E   &640   \\
$008.0+03.9$ & 180        & 20  & 34.6  & C G     &250   &$107.8+02.3$  & 280        & 25   & 16.4    & A E   &305   \\
$008.3-01.1$ & 90         & 20  & 9.9   & C H     &95    &$120.0+09.8$  & 420        & 70   & 27.6    & A E   &530   \\
$009.4-05.0$ & 130        & 30  & 12.8  & C H     &135   &$130.9-10.5$  & 140        & 40   & 59.3    & A E   &310   \\
$009.6+14.8$ & 270        & 60  & 14.3  & C H     &290   &$138.8+02.8$  & $<$ 150    & 50   & 23.1    & A E   &      \\
$010.1+00.7$ & 400        & 60  & 11.8  & C F H   &420   &$144.5+06.5$  & 215        & 30   & 34.9    & A E   &305   \\
$010.8-01.8$ & 130        & 20  & 11.9  & C H     &135   &$165.5-15.2$  & 220        & 30   & 94.4    & A E   &890   \\
$011.7-00.6$ & $<$ 150    & 50  & 8.7   & C H     &      &$166.1+10.4$  & 100        & 30   & 9.1     & A E   &105   \\
$020.9-01.1$ & 420        & 90  & 13.5  & C       &445   &$173.7+02.7$  & 260        & 20   & 11.5    & E     &270   \\
$025.8-17.9$ & 270        & 30  & 14.7  & C H     &290   &$194.2+02.5$  & 90         & 20   & 0.0     & A E H &90    \\
$027.7+00.7$ & 110        & 20  & 0.0   & B G     &110   &$197.8+17.3$  & 240        & 50   & 22.2    & A E H &280   \\
$033.8-02.6$ & 290        & 50  & 14.7  & E H     &310   &$206.4-40.5$  & 160        & 20   & 20.6    & C H   &180   \\
$034.6+11.8$ & 1220       & 100 & 11.9  & A E H   &1280  &$215.2-24.2$  & 1100       & 100  & 0.0     & C H   &1100  \\
$035.1-00.7$ & 170        & 20  & 13.0  & A E     &180   &$221.3-12.3$  & 350        & 50   & 11.0    & C H   &365   \\
$037.7-34.5$ & 375        & 40  & 14.9  & C H     &400   &$234.8+02.4$  & 350        & 20   & 16.5    & C H   &380   \\
$039.8+02.1$ & 240        & 20  & 0.0   & A E     &240   &$254.6+00.2$  & 1770       & 290  & 41.2    & F     &2800  \\ 
$041.8-02.9$ & 230        & 20  & 78.4  & A E H   &715   &$258.1-00.3$  & 180        & 40   & 4.5     & D H   &181   \\
$043.1+37.7$ & 230        & 30  & 10.8  & A E H   &240   &$259.1+00.9$  & 270        & 60   & 42.2    & D     &435   \\
$045.7-04.5$ & 140        & 20  & 25.4  & A B E H &170   &$261.0+32.0$  & 290        & 30   & 19.5    & C H   &330   \\
$050.1+03.3$ & 140        & 30  & 41.4  & A E H   &220   &$294.1+43.6$  & 130        & 20   & 47.3    & C H   &230   \\
$054.1-12.1$ & 130        & 40  & 8.9   & A E H   &135   &$342.1+10.8$  & $<$ 210    & 70   & 32.9    & H     &      \\
$063.1+13.9$ & 255        & 75  & 48.2  & A E     &460   &$349.5+01.0$  & 2150       & 220  & 14.8    & D F H &2310  \\
$064.7+05.0$ & 565        & 20  & 10.5  & A E     &585   &$352.6+00.1$  & 815        & 125  & 10.3    & F     &845   \\
$082.1+07.0$ & 250        & 50  & 11.0  & A E     &260   &$352.8-00.2$  & 420        & 50   & 14.8    & F     &450   \\
$083.5+12.7$ & 320        & 40  & 15.9  & A E     &350   &$358.5+02.6$  & $<$ 120    & 40   & 21.1    & C G   &      \\
$086.5-08.8$ & $<$ 120    & 40  & 10.4  & A E     &      &$358.5+05.4$  & 380        & 40   & 0.0     & C G   &380   \\
$089.0+00.3$ & 220        & 30  & 11.9  & A E     &230   &$359.3-00.9$  & 290        & 70   & 0.0     & F H   &290   \\
\hline                                           
\end{tabular}                                  
\begin{list}{}{}
\item[$^{\ast}$] References for 5 GHz measurements:
A) Gregory et al. (\cite{g_etal96}); B) Griffith et al. (\cite{g_etal95}); C) Griffith et al. (\cite{g_etal94}); D) Wright et al. (\cite{w_etal94});
E) Becker et al. (\cite{b_etal91}); F) Haynes et al. (\cite{h_etal79}); G) Milne (\cite{m79}); H) Milne (\cite{ma75})
\end{list}
\end{center}
\end{table*}

\section{The 43 Ghz Noto Survey}
\subsection{Sample Selection}
\label{s_s}
Our sample has been selected mainly from Condon \& Kaplan~(\cite{ck98}), who performed a cross-correlation between
the Strasbourg-ESO Catalogue of Galactic Planetary Nebulae (Cat. $<$V/84$>$, Acker et al.~\cite{a_etal92}) 
and the 1.4 GHz NRAO VLA Sky Survey (NVSS).
Among  these sources we selected only those whose flux densities at
1.4 GHz (NVSS) is higher than 100 mJy, for a total of 64 PNe.
In the conservative hypothesis of optically thin nebula at 1.4 GHz, a cut-off at $100$~mJy
guarantees a 43 GHz flux density easily detectable with the Noto Radiotelescope.
This estimated 43~GHz flux density will be a lower limit
in the case of optically thick nebula at 1.4 GHz.  
We also considered Condon et al.~(\cite{c_etal99}) who selected infrared PNs in NVSS by cross-correlation between NVSS and a sample 
from IRAS PSC on the basis of infrared colours characteristic of PNe.
Only 3 over the 122 of infrared PNs candidate have $S_{\mathrm {1.4~GHz}}\geq 100$ mJy, yielding to a {\it final} sample of 67
objects.

To avoid problems due to possible contamination in the Noto beam (HPBW$= 54{\arcsec}$), for each source of the sample, we have extract from NVSS 
a $25^{\prime}\times25^{\prime}$ field, centred at the position of the  target.
Sources that show a very extended ($\geq 2^{\prime}$)
emission  or that are  located in very high confusion region have been rejected.
This reduces our sample to 62 objects.
The list of the selected targets, with names and positions, is reported in Table~\ref{tab_posizioni}.

\subsection{Observations and Results}
In the last few years the INAF-IRA 32m Noto radiotelescope has been subjected to a 
series of structural improvements which have increased remarkably its potential capabilities 
as single dish instrument. In particular, the installation of an active surface allows to operate with good  
performances at high frequencies.
 
The observations reported in this paper were carried out in different epochs, between 2005 and 2006.
The 43 GHz supereterodyne receiver is cooled double polarization receiver, with typical zenith system temperature ($T_{\mathrm {sys}}$), in both channels, of the order of 80 -- 100 K, depending on the weather conditions. The gain ranges from 0.05 to 0.1 K/Jy (Leto et al. \cite{l06}), depending on elevation, and this determines a zenith System Equivalent Flux Density (SEFD) of 1600 -- 2000 Jy. 
The observations were performed with a 400 MHz instantaneous band.

All the sources of our sample were observed with the on the fly (OTF) scan technique, which consists in driving the beam of the telescope across the source in RA direction. The typical scan duration was of the order of 
20 secs, short enough to remain close enough to the white noise regime of the radiometer.
In order to achieve a good signal to noise ratio, each source was observed many times, for a
total integration time of 30 minutes. Multiple OTF scans were then added together reaching a typical rms of 2 -- 3 mK.

Daily gain curves were obtained and the flux scale was fixed by using NGC7027 as primary calibrator and 3C286 as secondary calibrator. 
The adopted flux densities for NGC7027 and 3C286 are respectively:
5.07 and 1.86 Jy. 
The assumed flux densities  are those reported by Ott et al.~(\cite{o_etal94}); in the case of NGC7027 the measure has been corrected
for the observed decreasing of 0.15 percent/year as reported by Perley et al.~(\cite{p_etal06}).


We detected 55 out of the 62 observed objects, with a 89 $\%$ detection rate. 
Results of such observations are reported in Table~\ref{tab_misure}, where the 43~GHz measured flux density, or its 3$\sigma$ upper limit, is listed.
In the 4$^{\mathrm {th}}$ and 10$^{\mathrm {th}}$ columns of the same table we report the angular size at 1.4 GHz of each source ($\theta_\mathrm{1.4 ~GHz}$), 
as derived from the analysis of the relative  NVSS map, which has an angular resolution of 45${\arcsec}$. 
The source angular size
has been derived as the geometrical mean of the minor and major axes obtained by fitting a two dimensional Gaussian 
at the source position in the map of the
brightness distribution, by using the task JMFIT of the NRAO {\bf A}stronomical {\bf I}mage 
{\bf P}rocessing {\bf S}ystem (AIPS).

To take into account for possible partial resolution of the source by the Noto beam, 
assuming that the  source size at 43 GHz is the same as measured at 1.4 GHz,
we corrected the measured flux density as 
\begin{displaymath}
S_\mathrm{c~43 ~GHz}=S_\mathrm{43 ~GHz} \times \frac{\theta _{\mathrm {Noto}} ^2 + \theta _{\mathrm {1.4 ~GHz}} ^2}{\theta _{\mathrm {Noto}} ^2}
\end{displaymath}
where $\theta _{\mathrm {1.4 ~GHz}}$ and $\theta _{\mathrm {Noto}}$ are the width of the source at 1.4 GHz (reported in Table~\ref{tab_misure})
and the width of the Noto radiotelescope beam (HPBW=$54\arcsec$) respectively.
Sources having angular size greather than 25${\arcsec}$, comparable to the 43 GHz Noto beam (HPBW), have flux corrections significantly larger than the rms associated to the flux density. This means that they cannot be regarded as point-like. 
The resulting fluxes $S_\mathrm{c~43 ~GHz}$ are listed in Table~\ref{tab_misure}.

\section{The SEDs}

\subsection{The free-free contribution}
\label{f_c}
As a first step in our analysis, we derived the spectral index $\alpha$,
($S_{\nu} \propto \nu^{\alpha}$), by combining the fluxes measured at 43 GHz with the 5 GHz single dish measurements available from literature. In order to  prevent any error due to resolving out some of the flux density, we considered only the objects
that have angular size, at 1.4 GHz ($\theta_\mathrm{1.4 ~GHz}$), lower than $25 \arcsec$. This reduces our sample 
 detected sources to 42 PNe that 
can be, quite confidentially, considered point-like with respect to the telescope beam at the observational frequency.
Literature 5 GHz data are available only for 41 targets. 
 The 5 GHz single dish measurements are from surveys performed by NRAO Green Bank 91 cm telescope, with a beamwidth (HPBW) $\approx 3.5{^\prime}$ (Gregory et al. \cite{g_etal96}, Becker et al. \cite{b_etal91}) and Parkers  64 m telescope
with a beamwidth $\approx 4.5^{\prime}$
(Griffith et al. \cite{g_etal95}, Griffith et al. \cite{g_etal94}, Wright et al. \cite{w_etal94}, Haynes et al. \cite{h_etal79}, Milne \cite{m79} and Milme \& Aller \cite{ma75}), as indicated in Table~\ref{tab_misure}.
For the sources observed in more than one survey, we
used the average 
of the fluxes from the different measurements.

\begin{figure}
\resizebox{\hsize}{!}{\includegraphics{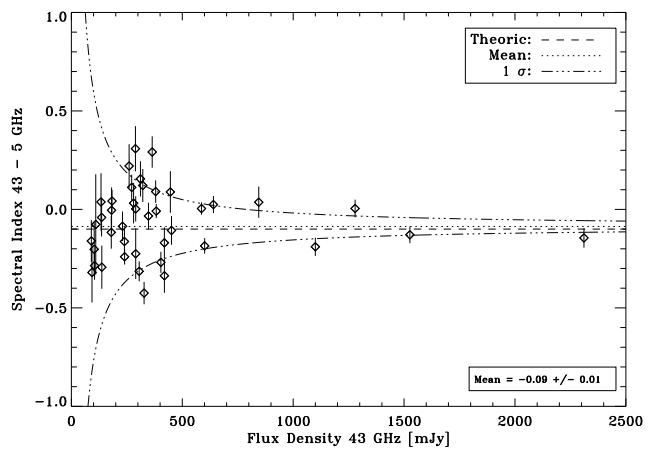}}
\caption{Calculated spectral index as function of flux at 43 GHz. As expected major dispersion is evident at lower fluxes}
\label{ind_spec}
\end{figure}

The spectral indices $\alpha$, calculated between the two frequencies, 
are shown in Fig.~\ref{ind_spec} as a function of the flux density at 43 GHz. 
Since its accuracy depends on the value of the measured flux densities and associated uncertainties,
we computed the expected $1~\sigma$ as a function of
the flux density, assuming a typical rms, associated to the 43~GHz and 5~GHz measurements, of 
$\sigma_{\mathrm{43 GHz}}\approx 50$~mJy and a $\sigma_{\mathrm{5 GHz}}\approx 35$~mJy, respectively. 
The $\pm 1~\sigma$ uncertainty values around the mean value $\alpha=-0.09$ 
are shown in Fig.~\ref{ind_spec} as dot-dashed lines. 
For about 70\% of the sources we get a spectral index inside the uncertainty lines,
showing that it is statistically consistent
with the value $\alpha=-0.1$, typical for an optically thin free-free emission.
We may conclude that for our sample, at least up to 43 GHz, the SED is
dominated by the free-free emission and other contributions,
if present, are negligible.

\subsection{The dust contribution}
\label{d_c}

PNs are usually surrounded by a dusty envelope which is the remnant of the precursor's wind.
Dust thermally re-radiates the UV radiation of the central star determining an excess that may extend from the far-IR to the radio region. Such a contribution is typically of the order of $40 \%$ of the total flux from a PNe (Zhang \& Kwok \cite{zk91}),
being more important in young PNe, since, in more evolved PNe, the circumstellar material has already dispersed.
We built the dust component of our sample of PNe by using IRAS measurements.

The IRAS data have been fitted by assuming that the
dust emits as a blackbody modified by the frequency dependent dust opacity, 
that is $F_{\nu}\propto \nu^p B_{\nu}(T_\mathrm{d})$,
where $B_\nu(T_\mathrm{d})$ is the Planck function for dust temperature $T_\mathrm{d}$,
and $p$ is the emissivity index. The index $p$  strongly depends on the mineralogical composition
of the grain and on their physical shape. 
In our calculation, following the results from detailed modelling of dust emission from PNe (Hoare \cite{hoare_90}), we assumed, for the slope of the far-IR emissivity law, a typical value of $p$= 1.5. Because strong contamination by line emission of $12~\mu$m IRAS band (Stasinska \& Szczerba \cite{ss99}) only 25, 60 and 100 $\mu$m fluxes have been used in the fitting procedure.

\begin{figure}
\resizebox{\hsize}{!}{\includegraphics{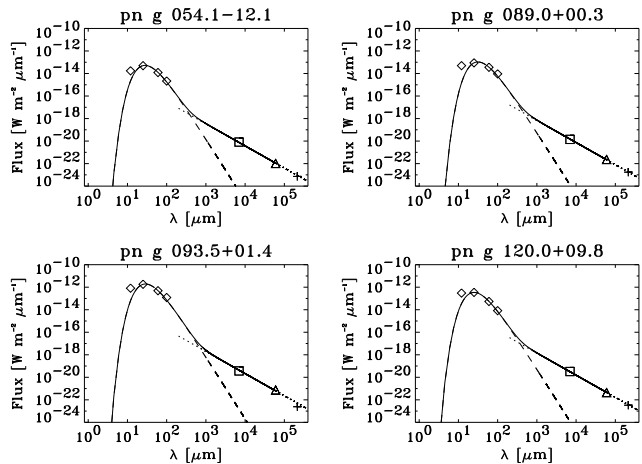}}
\caption{Spectral Energy Distributions of four of our targets. Both ionized gas (dotted line) and thermal dust (dashed line) contributions are shown.
Continuous line is the total emission. 
 Diamonds: IRAS data; squares: our measures at 43 GHz ($7\times 10^3 ~\mu$m); triangle: literature measures at 5 GHz ($6\times 10^4 ~\mu$m);
cross: NVSS data at 1.4 GHz ($2\times 10^5 ~\mu$m)}
\label{4sed}
\end{figure}

\section{Estimates of flux density at Planck channels}
The SEDs with both ionized gas  and dust contribution, evaluated in Sect.~\ref{f_c} and \ref{d_c}, were built for all the targets in our sample. In Fig.~\ref{4sed}, four of such SEDs are shown as an example.
In the SEDs, between radio and far-IR there is a large gap where measurements are missing.
New millimetre and sub-millimetre observations are clearly necessary  to better characterize and constraint the 
emission from different components, where the radio flux, from the ionized fraction of the nebula, could be merged to the dust contribution.  
This gap will be partially filled by the ESA-PLANCK mission, as the satellite is equipped with a Low Frequency Instruments (LFI), 
operating at 30, 44 and 70 GHz, and a High Frequency Instruments (HFI), operating at  100, 143, 217, 353, 545 and 857 GHz.

In order to evaluate the possibility that PLANCK will actually  detect sources in our sample, we should  compare, for each observational band, the expected flux from the sources to the foreseen sensitivity. 
The  expected fluxes are derived by summing, at each PLANCK channel,  both dust and ionized gas contribution. The first has been obtained by extrapolating the  modified blackbody fit (dashed lines in Fig.~\ref{4sed}); the latter has been obtained from fitting free-free emission 
from an optically thin nebula to the 5 GHz ($6\times 10^4 ~\mu$m) and 43 GHz ($7\times 10^3 ~\mu$m) 
data points and then extrapolating  to the Planck channels (dotted line in Fig.~\ref{4sed}),
 the NVSS measures at 1.4 GHz ($2\times 10^5 ~\mu$m) has been also displayed.

\begin{figure}
\resizebox{\hsize}{!}{\includegraphics{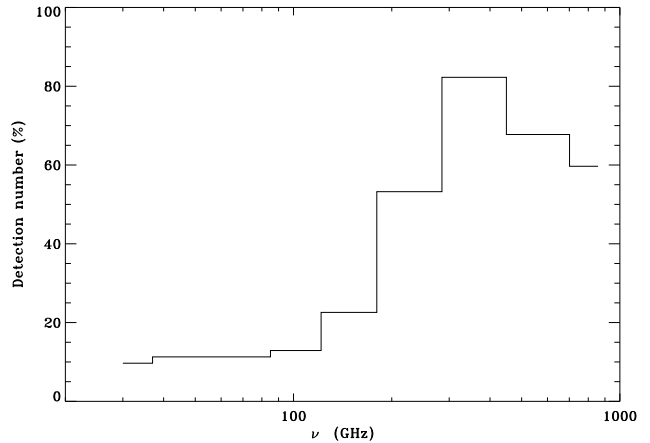}}
\caption{Histogram showing the fraction of planetary nebulae belonging to our sample that can be
detected by the Planck satellite}
\label{sens_plank}
\end{figure}

Following Umana et al. (\cite{Umana06}), the foreseen total sensitivity is assumed  to be the sum in quadrature of all the sources of confusion noise, namely:
the nominal instrumental Planck sensitivity per resolution element;
the Galactic and extragalactic foregrounds confusion noise (Toffolatti et al. \cite{toffolatti});
the CMB confusion noise (Bennett et al. \cite{bennett}).
All the considered confusion sources will contribute to a total sensitivity ($\sigma$) ranging from 270~mJy for the 30~GHz channel to 220~mJy for the 857~GHz channel.
In Fig.~\ref{sens_plank} we show the percentage of the objects from our sample that could be 
detected over the 3$\sigma$ threshold for each channels of the PLANCK instruments.
A small number of PNe belonging to our sample will be detected by PLANCK in the radio (LFI channels), while most of our targets will be easily seen by PLANCK in the millimetre and sub-millimetre bands (HFI channels). We may conclude that we foresee an important contribution of the PLANCK mission also to the study of PNe, as it would provide measurements in a very important, but still far to be fully explored, spectral region.

\section{Summary}

We have presented new 7 mm (43 GHz) observations of a sample of radio-bright PNe, carried out with the INAF-IRA Noto Radiotelescope. Such observations have been used to derive the high-frequency free-free contribution, due to the ionized fraction of the circumstellar envelope. 
So far, the majority of high frequency radio observations of PNe have been conducted with interferometers, that reveal details of source radio morphology but could, in principle, resolve out some of the extended emission.
Our single dish measurements provide an extended database of millimetre observations of PNe, to be used when is necessary to know the overall emission, such as when building a SED. 
We used our measurements, that trace the free-free contribution, together with IRAS measurements, which trace the thermal dust emission, to build up the observed SED, from radio to far-IR and by extrapolation, to estimate the expected fluxes in the 
spectral range between radio and sub-mm, where observational data are missing.

When comparing the expected millimetre-sub-millimetre fluxes with the total foreseen sensitivity of the forthcoming ESA mission PLANCK we estimate that a consistent number of our targets will be easily detected by PLANCK, mostly at higher frequency channels.  Therefore, even if designed for cosmological study, PLANCK could also contribute to the PNe science. 
Results from such kinds of observations, once modelled with appropriate code (i.e. CLOUDY, DUSTY), would provide important constraints on the chemical composition and structure of the CSEs. It would  point out the presence of multi-shells, related to multiple mass-loss event suffered from the central object during its previous evolution (AGB), or a contribution of different emission processes, besides free-free and thermal from dust, as recently claimed by Cassassus et al.~(\cite{cas_etal07}).
Moreover, PLANCK results can be considered as pathfinder for other future instrumentations, with the same frequency coverage, such as ALMA, as they will help in planning more focused experiments aimed to investigate the morphological details of the sources.

\begin{acknowledgements}
Based on obervations with the Noto Telescope operated by INAF-Istituto di Radioastronomia
This work has been partially supported by ASI through contract Planck LFI Activity of Phase E2.  
\end{acknowledgements}
{}

\end{document}